\documentstyle[prl,twocolumn,aps,epsf]{revtex}
\begin{document}
\title { Small size boundary effects on two-pion interferometry }
\author{Q.H. Zhang and Sandra S. Padula}
\address{Instituto de F\'\i sica Te\'orica, Rua Pamplona 145, 01405-900 
S\~ao Paulo, Brazil}

%\date{\today}
\maketitle
\begin{abstract}
Bose-Einstein correlations of 
two identically charged pions are derived when these particles, 
the most abundantly produced in relativistic heavy ion collisions, are 
confined in finite volumes. Boundary effects on single-pion 
spectrum are also studied. Numerical 
results emphasize that conventional formulation usually adopted to 
describe two-pion interferometry should not be used when the source size 
is small, since this is the most sensitive case to boundary effects. 
Specific examples are considered for better illustration.  
\end{abstract}
\pacs{25.75.-q, 25.75.Gz, 25.70.Pq}
\vskip -1.35cm
%{\it PACS number(s): 25.75.-q, 25.75.Gz, 25.70.Pq}

\section{Introduction}

	It is generally expected that high energy heavy-ion collisions 
may provide the tools to probe the existence of a new phase of matter of 
strongly interaction particles, the quark-gluon plasma (QGP), at high
temperature and high baryon density\cite{Wong1}. The hope of discovering 
the QGP in high energy heavy-ion collisions is to some extent connected 
to the possibility of measuring the geometrical sizes of the emission region 
of secondary particles. An important tool for accomplishing such 
size measurements is the so-called Hanbury-Brown-Twiss (HBT)
interferometry\cite{HBT}. This method was 
originally proposed in the 50's for measuring stellar radii but, shortly  
afterwards, it was discovered\cite{GGLP} that a similar procedure could 
also be applied to high energy collisions for determining the dimensions 
of pion emitting sources. This method has extensively been developed,  
improved, and better understood since the pioneering times\cite{boal}.

\bigskip
	Differently from the stellar case, however, where the dimensions 
are indeed immense, in the subatomic level the effects associated to the 
small sizes of the particle emitters and their boundaries may have an 
important role. Indeed, already in the well-known paper by  
Gyulassy, Kauffmann, and Wilson\cite{GKW79}, and more recently, in 
Ref.\cite{weiner,MW93,YS94,MW95,AS97,YS99}, effects of source 
finiteness on particle spectra and correlation functions were considered, 
although the conclusions of some of them were somewhat contradictory. 
For example, the low transverse momentum region of Ref.\cite{YS94,YS99} 
is shown to be enhanced with respect to the Bose-Einstein distribution. 
However, this enhancement was not observed in other references quoted 
above. As we shall see later, in agreement with results of 
Ref.\cite{MW93,MW95,AS97}, a depletion in the low momentum region is 
observed instead. This apparent discrepancy may be explained by both 
the form chosen for the density matrix in Ref.\cite{YS94,YS99}, 
and by the full field theoretical approach adopted there. However, 
the inherent difficulties of that approach are enormous, and the simpler 
treatment discussed in the present paper already sheds light to 
the relevant points of the problem.  

\bigskip
The approach suggested in Ref.\cite{MW93,MW95} seemed 
appealing for the following reasons. First, it considers that 
in an ultra-relativistic nucleus-nucleus collisions pions 
are the most abundant produced particles, being emitted at 
freeze-out temperatures  around 0.1-0.2 GeV.
Following Ref.\cite{MW95,AS97}, it is argued that right after these 
collisions, since the average pion separation is smaller than their 
interaction range, the pions in such a stage of the 
system evolution are still strongly interacting with each other. 
The effects of interaction among pions could 
then be modeled by considering that they are moving in an attractive 
mean field potential, which extends over the whole pion system. 
This implies, for instance, that  
in the two-pion case, they would not suffer any other effects besides 
the mean field attraction and the identical particle symmetrization. 
Consequently, rather than being in a gas, the pion system should 
be considered in a quasi-bound liquid phase, with the surface 
tension\cite{shu} acting as a reflecting boundary. Although details on 
this reflection depend on the pion wavelength, the pion wave function 
could be considered as vanishing outside this boundary. 
The pions become free when their average separation is larger 
than their interaction range. Due to the short range of the 
strong interaction, however, we would expect this liquid-gas 
transition to happen very rapidly, in such a way that  
the momentum distribution of pions could be essentially governed by their 
momentum distribution just before they freeze out. Under these circumstances, 
we would also expect that the observed pion momentum distribution would be 
modified by the presence of this boundary. And this is, in fact, what is 
analyzed in this work, as well as in Ref.\cite{MW93,YS94,MW95,AS97,YS99}.

\bigskip
On the other hand, since pion interferometry is sensitive to the geometrical 
size of the emission region as well as to the underlying dynamics, we would 
expect that the boundary would also affect the correlation function, but 
a priori we would not know  how. Would it affect single- and double 
inclusive distributions similarly? How would be intercept of the 
two-particle correlation function behave? How would the general 
shape of this function be affected? For an insight into these 
questions, we here investigate 
the effects exerted by the boundary on the two-particle correlation
function. We could naively expect that the importance of 
quantum statistics would progressively increase as the dimension of 
the emission region decreases. The results turned to exactly 
fulfill these expectations. Consequently, semi-classical 
approaches would have their applicability limited by the size of the 
emission region in focus. In other words, small emission volumes 
would stress the need for quantum statistics and, as a consequence, 
classical density matrices would lead to inconsistent results. This 
problem is clearly illustrated later in the present work. 

\bigskip
The plan of this paper is the following:  in section II, we  
derive the single-inclusive distribution, as well as the two-pion 
correlation function, considering a density matrix suited for 
describing $\pi^{\pm} \pi^{\pm}$ Bose-Einstein effects. 
In section III, the boundary effects on the two-pion correlation and single 
particle spectrum distribution are illustrated 
by means of two specific examples. Section IV is devoted to illustrate 
the results which would be expected when previous 
methods for deriving two-pion interferometry formula are employed  
and the finite volume effects are studied. Finally, conclusions are 
discussed in section V. 

\section{Spectrum and Two-pion correlation function}

In this section, we derive a generic formulation for the single-
as well as for the two-particle inclusive distributions, which 
would be suited for describing $\pi^+ \pi^+$ or $\pi^- \pi^-$ 
bounded in a finite volume.

We assume the pion creation operator in coordinate space 
can be expressed as 
\begin{eqnarray}
\hat{\psi}^{\dagger}({\bf x})=
\sum_{\lambda}\hat{a}^{\dagger}_{\lambda}\psi^*_{\lambda}({\bf x}), 
\end{eqnarray}
where $a^{\dagger}_{\lambda}$ is the creation operator for creating a 
pion in a quantum state characterized by a 
quantum number $\lambda$. Then, 
$\psi_{\lambda}({\bf x})$ is one of eigenfunctions belonging to  
a localized complete set, which 
satisfies the orthonormality condition 
\begin{equation}
\int d{\bf x} \psi^*_{\lambda}({\bf x})
\psi_{\lambda'}({\bf x})=\delta_{\lambda,\lambda'},
\end{equation}
and completeness relation
\begin{equation}
\sum_{\lambda}\psi^*_{\lambda}({\bf x})\psi_{\lambda}({\bf y})=
\delta({\bf x}-{\bf y}).
\label{xx3}
\end{equation}

Similarly, the pion annihilation operator in coordinate space can be 
written as
\begin{equation}
\hat{\psi}({\bf x})=\sum_{\lambda}\hat{a}_{\lambda}\psi_{\lambda}({\bf x}). 
\end{equation}

In momentum space, the corresponding pion creation 
operator, $\hat{\psi}^{\dagger}({\bf p})$, and 
annihilation operator, $\hat{\psi}({\bf p})$, 
can be expressed as
\begin{equation}
\hat{\psi}^{\dagger}({\bf p})=\sum_{\lambda} \hat{a}^{\dagger}_{\lambda} 
\tilde{\psi}^*_{\lambda}({\bf p})
\end{equation}
and 
\begin{equation}
\hat{\psi}({\bf p})=\sum_{\lambda} \hat{a}_{\lambda} 
\tilde{\psi}_{\lambda}({\bf p}) , 
\end{equation}
where
\begin{equation}
\tilde{\psi}_{\lambda}({\bf p})=\frac{1}{(2\pi)^{3/2}}
\int \psi_{\lambda}({\bf x})e^{i{\bf p}\cdot {\bf x}}d{\bf x}   .
\label{Fourier}\end{equation}

We write the density matrix operator for our bosonic system  
as 

\begin{equation}
\hat{\rho}=\exp\left[-\frac{1}{T} (\hat{H}-\mu \hat{N})\right]
\; \;, \label{rho}\end{equation}
where 
\begin{equation}
\hat{H}=\sum_{\lambda}E_{\lambda}a^{\dagger}_{\lambda}a_{\lambda}, \; \;
\hat{N}=\sum_{\lambda}a^{\dagger}_{\lambda}a_{\lambda} 
\; \;, \label{HN}\end{equation}
are the Hamiltonian and number operators, respectively; T is the temperature. 
 
 	The corresponding normalization is explicitly included 
in the definition of the expectation value of observables as, 
for instance, for an operator $\hat{A}$ 

\begin{equation}
\langle \hat{A}\rangle =\frac{tr\{ \hat{\rho} \hat{A}\}}{tr\{\hat{\rho}\}}
\; \;.\label{A}\end{equation}

 Then, the single-pion distribution can be 
written as
\begin{eqnarray}
P_1({\bf p})&=&\langle \hat{\psi}^{\dagger}({\bf p})\hat{\psi}({\bf p})\rangle 
\nonumber\\
&=&\sum_{\lambda}\sum_{\lambda'}
\tilde{\psi}_{\lambda}^*{(\bf p)}
\tilde{\psi}_{\lambda'}^*{(\bf p)}
\langle \hat{a}^{\dagger}_{\lambda}\hat{a}_{\lambda'}\rangle 
\; \;.\label{P1in}\end{eqnarray}

The expectation value 
$ \langle \hat{a}^{\dagger}_{\lambda}\hat{a}_{\lambda'} \rangle $  
is related to the occupation probability of the 
single-particle state $\lambda$,   
$N_\lambda$,  by 
\begin{equation}
\langle \hat{a}^{\dagger}_{\lambda}\hat{a}_{\lambda'} \rangle
=\delta_{\lambda,\lambda'}N_{\lambda}
\; \; ;\label{adagaa}\end{equation}
for a bosonic 
system in equilibrium at a temperature $T$ and chemical 
potential $\mu$, 
it is represented by the Bose-Einstein distribution 
\begin{equation}
N_{\lambda}=\frac{1}{\exp{\left[\frac{1}{T} (E_{\lambda}-\mu)\right]}-1 }
\; \;.\label{Nlamb}\end{equation}

By inserting Eq. (\ref{adagaa})  and (\ref{Nlamb}) into (\ref{P1in}), we 
obtain the single-particle spectrum for one pion species as
\begin{equation}
P_1({\bf p})=\sum_{\lambda}N_{\lambda}
\tilde{\psi}^*_{\lambda}({\bf p})
\tilde{\psi}_{\lambda}({\bf p})
\;.\label{p1g}\end{equation}

The above formula coincides with the one employed in Ref.\cite{MW95}  
for expressing the single-pion distribution. 

Similarly, the two-pion distribution function can be written as
\begin{eqnarray}
P_2({\bf p_1,p_2})&=&
\langle  \hat{\psi}^{\dagger}({\bf p_1})\hat{\psi}^{\dagger}({\bf p_2})
\hat{\psi}({\bf p_1})\hat{\psi}({\bf p_2})\rangle
\nonumber\\
&=&\sum_{\lambda_1,\lambda_2,\lambda_3,\lambda_4}
 \tilde{\psi}_{\lambda_1}^*({\bf p_1})
\tilde{\psi}_{\lambda_2}^*({\bf p_2})
\tilde{\psi}_{\lambda_3}({\bf p_1})
\tilde{\phi}_{\lambda_4}({\bf p_2})
\nonumber\\
&&
\langle  \hat{a}^{\dagger}_{\lambda_1}\hat{a}^{\dagger}_{\lambda_2}
\hat{a}_{\lambda_3}\hat{a}_{\lambda_4} \rangle
\nonumber\\
&=&\sum_{\lambda_1,\lambda_2,\lambda_3,\lambda_4}
\tilde{\psi}_{\lambda_1}^*({\bf p_1})
\tilde{\psi}_{\lambda_2}^*({\bf p_2})
\tilde{\psi}_{\lambda_3}({\bf p_1})
\tilde{\phi}_{\lambda_4}({\bf p_2})
\nonumber\\
&&
\left[\langle \hat{a}^{\dagger}_{\lambda_1}\hat{a}_{\lambda_3}\rangle 
\langle \hat{a}^{\dagger}_{\lambda_2}\hat{a}_{\lambda_4}\rangle_{\lambda_1 \ne \lambda_2} 
\right.
\nonumber\\
&&
+ \langle \hat{a}^{\dagger}_{\lambda_1}\hat{a}_{\lambda_4}\rangle
\langle \hat{a}^{\dagger}_{\lambda_2}\hat{a}_{\lambda_3}\rangle_{\lambda_1 \ne \lambda_2} 
\nonumber\\
&&
\left. +\langle \hat{a}^{\dagger}_{\lambda_1}\hat{a}^{\dagger}_{\lambda_2}
\hat{a}_{\lambda_3}\hat{a}_{\lambda_4}\rangle_{\lambda_1=\lambda_2=\lambda_3=\lambda_4}
\right ]
\nonumber\\
&&
=P_1({\bf p_1})P_1({\bf p_2})+ \sum_{\lambda_1}\sum_{\lambda_2}
\nonumber\\
&&
N_{\lambda_1}N_{\lambda_2}
\tilde{\psi}^*_{\lambda_1}({\bf p_1})
\tilde{\psi}_{\lambda_1}({\bf p_2})
\tilde{\psi}^*_{\lambda_2}({\bf p_1})
\tilde{\psi}_{\lambda_2}({\bf p_2})
\nonumber\\
&=&P_1({\bf p_1})P_1({\bf p_2})+
|\sum_{\lambda}N_{\lambda}\tilde{\psi}_{\lambda}^*({\bf p_1})
\tilde{\psi}_{\lambda}({\bf p_2})|^2
.\nonumber\\
\label{p2g}\end{eqnarray}

Since we are considering the case of two indistinguishable, 
identically charged pions, then 
\begin{equation}
\langle \hat{a}_{\lambda}^{\dagger}\hat{a}_{\lambda}^{\dagger}
\hat{a}_{\lambda}\hat{a}_{\lambda}\rangle =
2\langle \hat{a}^{\dagger}_{\lambda}\hat{a}_{\lambda}\rangle^2.
\label{BE1}
\end{equation}

	From the particular form proposed for 
the density matrix in Eq. (\ref{rho}), we can see that
$\langle \hat{a}^{\dagger}_{\lambda}\hat{a}^{\dagger}_{\lambda} \rangle  = 
 \langle \hat{a}_{\lambda}\hat{a}_{\lambda} \rangle = 0$, 
showing that it would not be suited for describing $\pi^0 \pi^0$ and 
$\pi^+ \pi^-$ cases. For this purpose, the formalism proposed in 
Ref.\cite{weiner,YS94,YS99} may be more adequate. 

The two-particle 
correlation can be written as
\begin{eqnarray}
C_2({\bf p_1,p_2})&=&\frac{P_2({\bf p_1,p_2})}{P_1({\bf p_1})
P_1({\bf p_2})}
\nonumber\\
&=&1+\frac{|\sum_{\lambda}N_{\lambda}
\tilde{\psi}_{\lambda}^*({\bf p_1})
\tilde{\psi}_{\lambda}({\bf p_2})|^2}
{\sum_{\lambda}N_{\lambda}|\tilde{\psi}_{\lambda}({\bf p_1})|^2
\sum_{\lambda}N_{\lambda}|\tilde{\psi}_{\lambda}({\bf p_2})|^2}
\; \;. \nonumber\\
\label{pi2}\end{eqnarray}

We can see immediately from the above formula that if ${\bf q}={\bf p_1}-
{\bf p_2}=0$ we have $C_2({\bf p,p})=2$. 
We also notice that the result for the correlation function 
in Eq. (\ref{pi2}) reflects the 
symmetrization over different states (and thus, the 
uncertainty in the determination of the pion state). 

Within this formulation we can also define the corresponding 
Wigner function, $g({\bf x},{\bf K})$, as
\begin{eqnarray}
g({\bf x},{\bf K})&=
&\frac{1}{(2\pi)^3}
\sum_{\lambda}N_{\lambda}\int \psi_{\lambda}^*\left({\bf x+\frac{y}{2}}\right)
\psi_{\lambda}\left({\bf x-\frac{y}{2}}\right)
\nonumber\\
&&
\exp(-i{\bf K}\cdot {\bf y}) d{\bf y},
\label{Wignerx}
\end{eqnarray}

	Consequently, we can write  
\begin{eqnarray}
\langle \hat{\psi}^{\dagger}({\bf p_1})\hat{\psi}({\bf p_2})\rangle
&=&\sum_{\lambda}N_{\lambda}\tilde{\psi}^*_{\lambda}({\bf p_1})
\tilde{\psi}_{\lambda}({\bf p_2})
\nonumber\\
&=&\int e^{-i({\bf p_1-p_2})\cdot {\bf x}}\; g({\bf x},{\bf k}) d{\bf x}.
\end{eqnarray}

By means of this Wigner function, the two-pion interferometry formula 
can be re-written as\cite{pratt,pgg,Heinz96}
\begin{eqnarray}
C_2({\bf p_1,p_2})&=& 1 +
\nonumber\\
&&
\frac{\int e^{i{\bf q}\cdot ({\bf x-y})} 
g({\bf x},{\bf K}) g({\bf y},{\bf K})
d{\bf x} d{\bf y}}
{\int g({\bf x},{\bf p_1}) g({\bf y},{\bf p_2})d{\bf x}d{\bf y}}.
\label{e1}
\end{eqnarray}

In the above equation, ${\bf K=(p_1+p_2)}/2$ is 
the two-pion average momentum, and ${\bf q=p_1-p_2}$ is 
their relative momentum.  Here $g({\bf x,K})$ can be 
interpreted as the probability of finding a
pion at point {\bf x} with momentum {\bf K}. 

\section{Two-pion correlation from a finite volume}

\subsection{Example 1}

In order to investigate the effect of the boundary on the single- 
and on the two-pion distribution functions, 
we assume that pions produced in high energy heavy-ion 
collisions are bounded in a sphere, just before freezing-out. 
In other words, their distribution functions are essentially 
the the ones they had while confined. 
The pion wave function should be determined by the solution of the 
Klein-Gordon equation 
\begin{equation}
\left [ \Delta+k^2\right ] 
\psi({\bf r})=0 
\; \;, \label{kg}\end{equation}
where $k^2=E^2-m^2$ is the momentum of the pion. On writing the above 
equation, we have assumed that the potential felt by the pion inside 
the sphere is zero, while outside it is infinite. The  
boundary condition to be respected by the solution is 
\begin{equation}
\psi({\bf r})|_{r=R}=0
\; \; ,\label{genbc}\end{equation}
where $R$ is the radius of the sphere at freeze-out  time. 

The normalized wave function corresponding to the solution of the 
above equation can easily be written as

\begin{eqnarray}
\psi_{klm}({\bf r})&=&
 \frac{1}{RJ_{l+\frac{3}{2}}(kR)} \sqrt{\frac{2}{r}} Y_{lm}(\theta,\phi)
 J_{l+\frac{1}{2}}(kr)~~~~ (r < R),
\nonumber\\
&=& 0 ~~~~~~~~~~~~~~~~~~~~~~~~~~~~~~~~~~~~~~~~~~~~~~~ (r \geq R).
\nonumber\\
\label{psiklm1}\end{eqnarray}

The momentum of the bounded particle, $k$, can be determined as the 
solution of the equation
\begin{equation}
J_{l+\frac{1}{2}}(kR)=0
\; \;. \label{jlbc}\end{equation}

Inserting Eq. (23) into Eq. (7), we can determine the Fourier transform of 
the confined solution of a pion inside the sphere, as a function of the 
momentum {\bf p}, as 

\begin{eqnarray}
&&\tilde{\psi}_{klm}({\bf p})=\sqrt{\frac{2}{p}} i^l Y_{lm}(\hat{p})
\frac{1}{RJ_{l+\frac{3}{2}}(kR)}
\left( \frac{R}{p^2-k^2} \right)
\nonumber\\
&&\left[ pJ_{l+\frac{3}{2}}(pR)J_{l+\frac{1}{2}}(kR) 
- kJ_{l+\frac{1}{2}}(pR) J_{l+\frac{3}{2}}(kR) \right]  
\nonumber\\
&&~~~~~~~~~~~~~~~~~~~~~~~~~~~~~~~~~~~~~~~~~  (p \ne k),
\nonumber\\\ 
&=&\sqrt{\frac{1}{2p}} \; i^l \; Y_{lm}(\hat{p}) \; RJ_{l+\frac{3}{2}}(kR)
\; ~\; ~~~~(p=k)
. \label{psiklm2}\end{eqnarray}

On deriving the above equation, we have made use 
the following integral equations
\begin{eqnarray}
&&\int_0^{R}r dr J_{l+1/2}(pr)J_{l+\frac{1}{2}}(kr)=\frac{R}{p^2-k^2}
\nonumber\\
&&
\left[\frac{}{} pJ_{l+\frac{3}{2}}(pR)J_{l+\frac{1}{2}}(kR) - 
kJ_{l+\frac{1}{2}}(pR)J_{l+\frac{3}{2}}(kR)\right],
\label{x1}\end{eqnarray} 

and

\begin{eqnarray}
\int_0^{R} r dr J_{l+\frac{1}{2}}(kr)J_{l+\frac{1}{2}}(kr)
&=&\frac{R^2}{2}J^2_{l+\frac{3}{2}}(kR)
\; \;. \label{x2}\end{eqnarray}

Besides, by imposing that the solution should vanish at the 
boundary, expressed by Eq. (\ref{jlbc}), it can be shown that 

\begin{eqnarray}
&&\lim_{p\rightarrow k}
\frac{pJ_{l+3/2}(pR)J_{l+1/2}(kR)-kJ_{l+\frac{1}{2}}(pR)J_{l+\frac{3}{2}}(kR)}
{p^2-k^2}
\nonumber\\
&&
=\frac{RJ^2_{l+\frac{3}{2}}(kR)}{2}
\; \;, \label{lim}\end{eqnarray}
i.e., $\tilde{\psi}_{klm}({\bf p})$ is a continuous function 
of $p$ at $p=k$. 

 Using the condition expressed by Eq. (\ref{jlbc}), 
Eq. (\ref{psiklm2}) can be further simplified as 

\begin{equation}
\tilde{\psi}_{klm}({\bf p})=\sqrt{\frac{2}{p}} i^l Y_{lm}(\hat{p})
\left[ \frac{-k}{p^2-k^2} \right] J_{l+\frac{1}{2}}(pR) 
\; \;. \label{psiklm3}\end{equation}

In terms of Eq. (\ref{x2}), the single-inclusive distribution function 
is given by 

\begin{eqnarray}
P_1({\bf p})&=&\sum_{klm}N_{klm}
\tilde{\psi}_{klm}^*({\bf p})\tilde{\psi}_{klm}({\bf p})
\nonumber\\
&=&\sum_{klm}\frac{1}{\exp{(\frac{E_{kl}-\mu}{T})}-1}
\sqrt{\frac{2}{p}} (-i)^l Y^*_{lm}(\hat{p})
\sqrt{\frac{2}{p}} i^l 
\nonumber\\
&&
Y_{lm}(\hat{p})
 \left[ \frac{-k\cdot J_{l+\frac{1}{2}}(pR)}
 {p^2-k^2} \right]
\left[ \frac{-k\cdot J_{l+\frac{1}{2}}(pR)}{p^2-k^2} \right]
\nonumber\\
&=&
\sum_{k,l}\frac{1}{\exp{(\frac{E_{kl}-\mu}{T})}-1}
\left(\frac{2l+1}{2\pi p}\right)
\left( \frac{kJ_{l+\frac{1}{2}}(pR)}{p^2-k^2} \right)^2,
\nonumber\\
\label{P1sph1}\end{eqnarray}

where  we have used that 
\begin{equation}
\sum_{m=-l}^{m=l}Y_{lm}^*(\hat{p}_1)Y_{lm}(\hat{p}_2)=
\left(\frac{2l+1}{4\pi}\right)
P_l(\hat{p}_1\cdot\hat{p}_2).
\end{equation}
                 
    Up to this point we have considered the pions confined in a sphere, 
which required the wave function to have a sharp change at $r = R$. 
However, as discussed in Ref.\cite{sarkar}, it 
could be more appropriate to consider a smoother boundary, by 
softening the potential felt by the pion at 
$r = R$. Unfortunately, this procedure would turn the problem into a very 
complex one \cite{sarkar}, and beyond the scope of this paper. 
Nevertheless, as a diffuse boundary would cause a gradual decrease to zero 
of the pion wave function, it could be simulated by taking the limit 
$R \rightarrow \infty$ \cite{sarkar} in Eq. (\ref{x1}), i.e., 

\begin{equation}
\int_0^{\infty}J_{l+\frac{1}{2}}(pr)J_{l+\frac{1}{2}}(kr)r dr=
\frac{1}{k}\delta(p-k)
\label{xxy}
\end{equation}

and 

\begin{equation}
J_{l+\frac{3}{2}}(kR)\rightarrow \sqrt{\frac{2}{\pi kR}}
\end{equation}

\begin{equation}
\tilde{\psi}_{klm}=i^{l}Y_{lm}(\hat{p})
\sqrt{\frac{\pi}{pkR}}\delta(p-k).
\label{xxy1}
\end{equation}

Then, by imposing the completeness relation, Eq. (\ref{xx3}), we 
can show that 
\begin{equation}
\sum_{\lambda}
\tilde{\psi}_{\lambda}^*({\bf p})
\tilde{\psi}_{\lambda}({\bf p})=\frac{V}{(2\pi)^3}
\end{equation}

With the $\delta$ function in Eq. (\ref{xxy}) and Eq. (\ref{xxy1}), 
the single particle spectrum, 
in the limit $R \rightarrow \infty$, is then written as 
\begin{equation}
P_1({\bf p})=\frac{1}{\exp{(\frac{E_p-\mu}{T})}-1}\left[\frac{V}{(2\pi)^3}
\right] \; \;, \label{P1Rinf1}\end{equation}
where $V=\frac{4\pi}{3}R^3$ is the volume of the sphere. We see 
from    Eq. (\ref{P1Rinf1}) that 
the ordinary Bose Einstein distribution is recovered in the limit of a 
very large volume.

For $p=0$ and $V=\frac{4\pi}{3}R^3$, Eq. (\ref{P1sph1}) becomes

\begin{equation}
P_1({\bf p})|_{p=0}=V\sum_{n=1}^{\infty}\frac{1}{\exp{(\frac{E_n-\mu}{T})}-1}
\left( \frac{3}{4\pi^5n^2} \right) 
\; \;, \label{P1pp0}\end{equation}

where
\begin{equation}
E_n=\sqrt{\left( \frac{n\pi}{R} \right)^2+m_{\pi}^2}
\; \;. \label{En}\end{equation}

In the limit $R \rightarrow \infty$, we have 
\begin{equation}
P_1({\bf p})=\left[\frac{V}{(2\pi)^3}\right]
\frac{1}{\exp(\frac{m_{\pi-\mu}}{T})-1}
\; \;, \label{P1Rinf2}\end{equation}
which is consistent with Eq. (\ref{P1Rinf1}).

From Eq. (\ref{P1pp0}), we see that the intercept of the 
spectrum depends on the value of the radius: as $R$ increases, 
$E_n$ becomes smaller, and the maximum value of this distribution, 
corresponding to $|{\bf p}| = 0$, becomes higher. 
In all numerical estimates considered in the present work we have fixed 
$\mu = 0$. In Figure 1,
the normalized single-particle distribution is plotted 
as a function of $|{\bf p}|$. We have chosen a discrete normalization, 
obtained by imposing ${\cal N} = \sum_{i=1}^{N} P_1(p_i)$, 
where N refers to the total number of bins in which the distribution 
is subdivided. 

We clearly see from Figure 1 that, due to the boundary effects, the  
maximum value of $|{\bf p}|$ in the spectrum decreases for decreasing 
volumes, being 
always smaller than the case corresponding to the $R \rightarrow \infty$ limit. 
The explanation for this behavior can be understood in terms 
of the uncertainty principle, i.e., as the volume of the system decreases, 
the uncertainty in the pion coordinate decreases accordingly, causing 
a large fluctuation in the pion momentum distribution. We should notice 
that this result coincides with the one obtained in Ref.\cite{MW93,MW95}, 
and is opposite to the results from Ref.\cite{YS94,YS99}. 

Regarding the spectrum, we could also inquire how would the 
freeze-out temperature affect it and how would the finite size effect 
compare with the $R \rightarrow \infty$ limit for different temperatures. 
This is illustrated in Figure 2. The curves there should be compared 
in groups of two: solid ($T=0.14$ GeV) and  dotted ($T=0.11$ GeV). 
For emphasizing the differences and similarities as 
$|{\bf p}|$ increases, we plot the difference between the two curves in 
each group, $\Delta P_1({\bf p})=P_1({\bf p})|_{R=3fm}-P_1({\bf p})|_{R=\infty}$, 
in Figure 3. 
We see that the lower the temperature, the 
bigger is the difference between the curves of each group, in 
the small momentum region.  
Decreasing the temperature has a similar 
effect on the spectrum as decreasing the radius: in both cases the 
fluctuations in the small region of the pion spectrum increases and 
the corresponding maximum is reduced. 
 In other words, the boundary effects are more significant 
when we deal with systems whose dimensions and temperatures are 
small. 

	We should observe that, except in Fig. 2 and 3 where the 
temperature dependence is studied, we have fixed $T=0.12$ GeV. The 
reason for this relies on Shuryak's arguments\cite{shu}, according 
to which for temperatures in the range 
$0.1 \leq T \leq 0.2$ GeV, the excited pionic matter would be better 
described as in a liquid phase inside a surface created by their 
mutual interaction. He added that, for $T \ge 0.15$ GeV, the influence of 
resonances become important but they are not included in the present 
study. Therefore we chose $T=0.12$ GeV, which is also of 
the order of the recent experimental freeze-out temperature estimated 
from both interferometry and spectra.

%From the definition (Eq. (\ref{rho})), we have 
Similarly, we can write for the expectation value of the product of 
two pion creation operators in momentum space 

\begin{eqnarray}
&&\langle \hat{\psi}^{\dagger}({\bf p_1})\hat{\psi}({\bf p_2})\rangle
=\sum_{klm}\frac{\tilde{\psi}^*_{klm}({\bf p_1})
\tilde{\psi}_{klm}({\bf p_2})}
{\exp{\left( \frac{E_{kl}-\mu}{T} \right)}-1}
%%&=&\sum_{klm}\frac{1}{\exp{\left( \frac{E_{kl}-\mu}{T} \right)}-1}
%%\tilde{\psi}^*_{klm}({\bf p_1})
%%\tilde{\psi}_{klm}({\bf p_2})
\nonumber\\
&=&
\sum_{klm}\frac{1}{\exp{\left( \frac{E_{kl}-\mu}{T} \right)}-1}  \times 
\nonumber\\
&&\sqrt{\frac{2}{p_1}}
(-i)^l Y_{lm}^*(\hat{p}_1) \left[\frac{-k}{p_1^2-k^2}\right]
J_{l+\frac{1}{2}}(p_1R)
\nonumber\\
&&\sqrt{\frac{2}{p_2}}
(i)^l Y_{lm}^*(\hat{p}_2)\left[\frac{-k}{p_2^2-k^2}\right]
J_{l+\frac{1}{2}}(p_2R)
\nonumber\\
&=&  \sum_{kl}\frac{1}{\exp{\left( \frac{E_{kl}-\mu}{T} \right)}-1}
\sqrt{\frac{4}{p_1p_2}} 
\nonumber\\
&&
\frac{k^2}{(p_1^2-k^2)(p_2^2-k^2)}J_{l+\frac{1}{2}}(p_1R) 
\nonumber\\
&&J_{l+\frac{1}{2}}(p_2R)
\left(\frac{2l+1}{4\pi}\right)P_l(\hat{p}_1\cdot\hat{p}_2)
\; \;. \label{psi*psi}\end{eqnarray}

The two-pion interferometry correlation function can then be estimated
by inserting the above expression into Eq. (\ref{pi2}).
In general, this function will depend on  
the angle between ${\bf p_1}$ and ${\bf p_2}$.
For the sake of simplicity, we will consider ${\bf p_1}$ parallel 
to ${\bf p_2}$, implying that 
$P_l(\hat{{\bf p}}_1\cdot\hat{{\bf p}}_2=\pm 1)=(\pm 1)^{l}$. 
The results for two-pion interferometry corresponding to 
different values of the pair average momentum ${\bf K=(p_1+p_2)}/2$ 
but fixed temperature are 
shown in Figure 4. We can see that, as the pair average momentum increases, 
the apparent source radius becomes bigger,  
which is an interesting behavior, if we compare to results corresponding to 
expanding systems. In this last case, the probed part of the system decreases
with increasing average momentum. Naturally, our present approach does not 
consider the effects of expansion and the enlargement of the system's 
apparent dimensions with increasing K, seen in Figure 4, has its origin 
in the strong sensitivity to the dynamical matrix. 
This can be better understood by observing the 
presence of the weight factor $N_\lambda$ in Eq.(\ref{pi2}), with $N_\lambda$ 
expressed in Eq.(\ref{Nlamb}). The increase of the average momentum reflects 
the increase in the individual momenta ${\bf p_1}$ and ${\bf p_2}$, which 
comes from larger values of the sum coefficient, $k$, in Eq.(\ref{pi2}). 
This has two opposite effects: the factors $1/(p_i^2-k^2)$ give 
larger contribution for $k \sim p_i$. However, bigger values of $k$ would 
also make the exponential factor (with $\mu = 0$) drop faster. 
Being so, by increasing 
K we are effectively including a larger number of $k$ coefficients that 
contributes to the sum in Eq.(\ref{pi2}), with decreasing weight 
$\sim \exp{(-E_{kl}/T)}$. The interference of 
these extra terms corresponding to larger $k$ with the terms already 
considered in the sum would make the correlation function drop faster, 
consequently becoming narrower. 
Alternatively, we could understand these results by noticing that 
pions with larger momentum come 
from larger quantum $\lambda$ states which, in turn, correspond to a 
smaller spread in coordinate space. As the weight factor in 
Eq. (\ref{pi2}) is of Bose-Einstein form, 
larger quantum states will give a smaller contribution 
to the source distribution, causing the effective 
source radius to appear larger. In order to confirm that the weight 
factor in Eq.(\ref{pi2}) is the responsible for the behavior observed in 
Fig. 4, let us consider the case in which we choose it to be a constant 
factor, for instance $N_{\lambda}=1$. 
This situation could be derived from the Bose-Einstein distribution 
form by considering $T \gg 1$, so that the two-pion interferometry 
results would become insensitive to the average momentum, due to the 
very large values of the temperature.
The numerical result 
corresponding to this case is also shown in Fig. 4 (narrower curve). On the 
other hand, with the 
help of the completeness relation, Eq.(\ref{xx3}), and of Eq.(\ref{Fourier}), 
by also assuming that the pions are confined in a sphere, it is 
straightforward to derive the following $K$ independent two-pion 
correlation function 
\begin{equation}
C_2(q)=1+\frac{9}{q^4R^6}[R\cos(qR)-\frac{\sin(qR)}{q}]^2
.\label{comp1}
\end{equation}

For completeness, we also include in Fig. 4 the curve based on 
Eq.(\ref{comp1}), which coincides with our numerically generated curve, 
cross checking the correctness of our numerical calculation. 

Figure 5 shows the two-pion correlation function for increasing values 
of the spherical radius, i.e., for enlarging volumes. From it, we can 
clearly see that, as the confining volume increases, the source 
radius derived from two-pion correlation also increases, as would 
be expected. 
                  
Again, as discussed previously for the spectrum, we could estimate the 
effect of a diffuse boundary on the two-pion correlation function 
by considering the limit $R \rightarrow \infty$. 
By inserting Eq. (\ref{xxy1}) into Eq. (\ref{psi*psi}), remembering 
that in this limit we can take $\sum_k \rightarrow \int dk$, and 
using the previous result for the spectrum in this limit, 
Eq. (\ref{P1Rinf1}), we finally obtain that 

\begin{center}
\begin{eqnarray}
\mbox{ $ C_2({\bf p_1,p_2})$ =  } \left\{
\begin{array}{l}
\mbox{1 ~ ~  ~  ~  ~ (${\bf p_1}$ $\ne$ ${\bf p_2}$)} \\
\mbox{2 ~ ~  ~  ~  ~ (${\bf p_1}$ = ${\bf p_2}$)} \\
%%%\mbox{2 ~ ~  ~  ~  ~ (\boldmath{$p_1$} = \boldmath{\bf $p_2$})} \\
\end{array} \right.
\end{eqnarray}
\end{center}
as would be expected. 

To conclude this section we should keep in mind that, if the system size 
is very small, it would be sensitive to the boundary effects 
even if we considered a diffuse boundary. On the contrary, if the system 
size is very large, we would not expect a significant effect in neither 
the sharp nor the diffuse boundary case\cite{sarkar}.

\subsection{Example 2}

In this subsection, we will study the sensitivity of spectrum and of the 
two-pion correlation function to the system boundaries, 
by considering the pion system inside a box of 
dimensions $L \times L \times L$. We choose first 
periodic boundary conditions. In this case, the eigenfunction 
can be written as

\begin{equation}
\psi_{\bf k}({\bf r})=\frac{1}{\sqrt{V}}\exp(-i{\bf k}\cdot {\bf r})
\; \;. \label{psik}\end{equation}

Here ${\bf k}$ is the quantum number which satisfies 
following constraint

\begin{center}
\begin{eqnarray}
\mbox{ $ k_i \cdot L = 2 \; n_i \; \pi $ } \rightarrow \left\{
\begin{array}{l}
\mbox{\boldmath{i=1,2,3}} \\
\mbox{$ n_i =0, \pm 1,\pm 2,...$ } \\
\end{array} \right.
\; \; \label{hah1}\end{eqnarray}
\end{center}

The corresponding Fourier transform $\tilde{\psi}_{\bf k}({\bf p})$ 
can be expressed as
\begin{eqnarray}
\tilde{\psi}_{\bf k}({\bf p})&=&\frac{1}{(2\pi)^{3/2}}\frac{8}{\sqrt{V}}
\left[\frac{\sin\left[(p_1-k_1)L/2\right]}{p_1-k_1}\right]
\nonumber\\
&&\left[\frac{\sin\left[(p_2-k_2)L/2\right]}{p_2-k_2}\right]
\left[\frac{\sin\left[(p_3-k_3)L/2\right]}{p_3-k_3}\right]
.\label{hah2}\end{eqnarray}
 
The single-particle and two-particle distributions follow 
from Eq. (\ref{p1g}) and (\ref{p2g}). 
We should notice that, in the limit $L\rightarrow 0$, and 
using  the condition (\ref{hah1}), we find the  contribution  of only 
one state (${\bf k =0 }$) to the two-pion correlation function, 
resulting in 
\begin{equation}
C({\bf p_1,p_2})=2 
\; \;. \nonumber\\
\end{equation}

On the other hand, if we take the limit of  
$V \rightarrow \infty$, Eq. (\ref{hah2}) becomes
\begin{equation}
\psi_{\bf k}({\bf p})=\frac{1}{\sqrt{V}}(2 \pi)^{3/2}
\delta({\bf p-k}).
\end{equation}

With the above form for $\psi_{\bf k}({\bf p})$ in the limit of 
very large volumes, we obtain for the correlation function 
\begin{center}
\begin{eqnarray}
\mbox{ $ C_2({\bf p_1,p_2})$ =  } \left\{
\begin{array}{l}
\mbox{1 ~ ~  ~  ~  ~ (${\bf p_1}$ $\ne$ ${\bf p_2}$)} \\
\mbox{2 ~ ~  ~  ~  ~ (${\bf p_1}$ = ${\bf p_2}$)} \\
\end{array} \right.
\end{eqnarray}
\end{center}

If, instead of the periodic boundary conditions, 
we consider that the pions are confined in the box, 
i.e., we assume the potential outside it is infinite, 
then two classes of solutions are possible:

\begin{equation}
\psi^I({\bf x})=\sqrt{\frac{8}{V}}\sin(k_1\cdot x) \sin(k_2 \cdot y)
\sin(k_3\cdot z)
\; \;, \label{psiI}\end{equation}

with
\begin{center}
\begin{eqnarray}
\mbox{ $ k_i \cdot L = 2 \; n_i \; \pi $ } \rightarrow \left\{
\begin{array}{l}
\mbox{\boldmath{i=1,2,3}} \\
\mbox{$ n_i = 1, 2,...$} \\
\end{array} \right.
\; \;, \label{kLIbc}\end{eqnarray}
\end{center}
and 
\begin{equation}
\psi^{II}({\bf x})=\sqrt{\frac{8}{V}}\cos(k_1\cdot x) \cos(k_2 \cdot y)
\cos(k_3\cdot z)
\; \;, \label{psiII}\end{equation}
with
\begin{center}
\begin{eqnarray}
\mbox{ $ k_i \cdot L = (2 n_i-1) \pi $ } \rightarrow \left\{
\begin{array}{l}
\mbox{\boldmath{i=1,2,3}} \\
\mbox{$ n_i = 1, 2,...$} \\
\end{array} \right.
\; \;. \label{kLIIbc}\end{eqnarray}
\end{center}

It can be shown that, for $V\rightarrow 0$,
we have 
\begin{equation}
C_2^{I,II}({\bf p_1,p_2})=2,
\label{c2Iv0II}\end{equation}
while, for $V \rightarrow \infty$, we obtain  
\begin{center}
\begin{eqnarray}
\mbox{ $ C_2^{I,II}({\bf p_1,p_2}) $ =  } \left\{
\begin{array}{l}
\mbox{1 ~ ~  ~  ~  ~ ~ ~(${\bf p_1} \ne \pm {\bf p_2}$)} \\
\mbox{2 ~ ~  ~  ~  ~ ~ ~(${\bf p_1}  =  \pm {\bf p_2}$)} \\
\end{array} \right.
\label{c2IvinfII}\end{eqnarray}
\end{center}

The reason for including the ($\pm$) signs in Eq.(\ref{c2IvinfII}) 
comes from the parity property of Eq. (\ref{psiI}) and (\ref{psiII}). 
From them, it is immediate to see that 
solution {\bf I} has negative parity, while solution {\bf II} has 
positive parity. 
The corresponding Fourier transforms then show the same parity property, 
i.e.,  

\begin{equation}
\tilde{\psi}^I_{{\bf k}}({\bf p})=-\tilde{\psi}^I_{{\bf k}}(-{\bf p}) 
\; \; \; ; \; \; \;  
\tilde{\psi}^{II}_{{\bf k}}(-{\bf p})=\tilde{\psi}^{II}_{{\bf k}}({\bf p})
\; \;. \label{parity}\end{equation}

From the above results, we can show that 
the single particle spectrum and two-pion correlation  function 
correspondingly have the following properties

\begin{equation}
P_1^{I,II}({\bf p})=P_1^{I,II}(-{\bf p}),
\end{equation}
and 
\begin{equation}
C_2^{I,II}({\bf p_1,-p_2})=C_2^{I,II}({\bf p_1,p_2})
\; \;. \label{sim1}
\label{c2par}\end{equation}

In particular, we see from Eq. (\ref{c2Iv0II}), (\ref{c2IvinfII}), 
and  (\ref{c2par}) that, if we choose ${\bf p_1=-p_2=p}$, we 
immediately get $C_2^{I,II}({\bf p_1=-p_2=p})=2$ for the 
confined boundary condition in both volume limits. 
That is the reason why, as a consequence 
of the parity property of the wave function, 
Eq. (\ref{c2IvinfII}) could be  
extended to $C_2^{I,II}({\bf p_1= \pm p_2= p})=2$. 

For periodical boundary condition, however, we have 
\begin{equation}
\tilde{\psi}_{\bf k}({\bf p})=\tilde{\psi}_{-{\bf k}}(-{\bf p}).
\end{equation}
Then, for the  single particle distribution we will have the  
following relation 

\begin{equation}
P_1({\bf p})=P_1(-{\bf p}).
\end{equation}

Nevertheless, the two-pion correlation function, 
which can be written as 
\begin{equation}
C_2({\bf p_1,-p_2})=
1+\frac{|\sum_{\bf k} N_{\bf k}
\tilde{\psi}_{\bf k}({\bf p_1})
\tilde{\psi}_{\bf k}(-{\bf p_2})|^2}
{\sum_{\bf k}N_{\bf k}|\tilde{\psi}_{\bf k}({\bf p_1})|^2
\sum_{\bf k}N_{\bf k}|\tilde{\psi}_{\bf k}({\bf p_2})|^2},
\end{equation}
in the case of periodical boundary condition,
will show no well-defined property under momentum reflection as 
the one expressed by Eq. (\ref{sim1}).

\section{Conventional HBT formulation}

We now discuss the case of the conventional formulation, usually 
adopted in HBT analysis, in terms of classical currents 
$j(x)$\cite{GKW79} representing the pion sources. 
For simplicity, we consider the  momentum as 
the only quantum number involved in the problem, i.e., we denote 
$\{\lambda\}$ as $\{ {\bf p}\}$. Besides, we also assume that the 
pion state could be characterized by the measured momentum. 
For instance, $\psi({\bf p_1})$ represents a 
pion in a quantum state denoted by ${\bf p_1}$.
Then the single-particle and the two particle distributions can be 
written as
\begin{equation}
P_1({\bf p})=N({\bf p})\psi^*({\bf p})\psi({\bf p}),
\end{equation}
and
\begin{equation}
P_2({\bf p_1,p_2})=N({\bf p_1})N({\bf p_2})
\psi^*({\bf p_1})\psi^*({\bf p_2})\psi({\bf p_1})\psi({\bf p_2}).
\end{equation}

In the above relations we dropped the subscript $\{\lambda\}$ of the 
state. $N({\bf p})$ is the 
Bose-Einstein distribution. In order to connect to the HBT effect, 
we need to make one further assumption: we assume that 
the source $j({\bf x})$ is chaotic and a function of the coordinates only. 
$\psi({\bf x})$ is determined as a solution of the equation
\begin{equation}
(\Delta + p^2)\psi^j(x)=j(x)
\;.\label{psij}\end{equation}

The superscript $j$ is introduced as a reminder that $\psi^j$ is the 
solution of equation (\ref{psij}), in the presence of the source $j(x)$. 
Then $\psi^j(x)$ can be written by 
\begin{eqnarray}
\psi^j(x)&=&\int G(x,x')j(x')d{\bf x'}
\nonumber\\
&=&
\frac{1}{(2\pi)^{3/2}}
\int e^{-i{\bf p}\cdot ({\bf x}-{\bf x'})} j({\bf x'}) d{\bf x'}.
\end{eqnarray}

In the above expression we have used the fact that $j(x')$ is localized in 
a small volume. The corresponding function in momentum space is then 
 
\begin{equation}
\psi^j({\bf p})=\int j(x)e^{ipx} d{\bf x}.
\end{equation}

The current $j(x)$ can be expressed as
\begin{equation}
j(x)=\sum_{i=1}^{N} A_i j_i(x),
\end{equation}
where $i$ denotes the number of the collision center; $A_i$ is a weight 
factor which represents the amplitude of the emitter. 
Assuming they are chaotic and a function of the coordinates only, then 
\begin{equation}
\{ j_i^*(x)j_j(y) \}=\delta_{ij} j_i^*(x)j_i(y).
\end{equation}

Here, $\{ \cdot\cdot\cdot \}$ denotes average over phases. 
The single particle spectrum and two-pion distribution function 
are then written as:
\begin{equation}
P_1({\bf p})=N({\bf p})\sum_i|A_i|^2 |j_i({\bf p})|^2,
\label{p1semi}\end{equation}
and
\begin{eqnarray}
P_2({\bf p_1,p_2})&=&N({\bf p_1})N({\bf p_2})
\nonumber\\
&&
[\sum_i|A_i|^2|j_i({\bf p_1})|^2
\sum_k|A_k|^2 |j_k({\bf p_2})|^2 +
\nonumber\\
&&
|\sum_{i=1}^N|A_i|^2  j_i^*({\bf p_1})j_i({\bf p_2})|^2
\nonumber\\
&&
-\sum_{i=1}^N |A_i|^4 j_i^*({\bf p_1})j_i^*({\bf p_2})j_i({\bf p_1})
j_i({\bf p_2})],
\label{p2semi}\end{eqnarray}
with

\begin{equation}
j_i({\bf p})=\frac{1}{(2\pi)^{3/2}}\int j_i(x)e^{ipx} dx.
\end{equation}

Inserting Eq. (\ref{p1semi}) and (\ref{p2semi}) into (\ref{pi2}), 
we obtain the two pion interferometry formula expressed as
\begin{eqnarray}
C_2({\bf p_1,p_2})&&=
1+\frac{|\sum_i |A_i|^2 j_i^*({\bf p_1})j_i({\bf p_2})|^2}
{\sum_i |A_i|^2 |j_i({\bf p_1})|^2
\sum_k |A_k|^2 |j_k({\bf p_2})|^2}
\nonumber\\
&&
-
\frac{\sum_i |A_i|^4 |j_i^*({\bf p_1})j_i({\bf p_2})|^2}
{\sum_i |A_i|^2 |j_i({\bf p_1})|^2
\sum_k |A_k|^2 |j_k({\bf p_2})|^2}
.\label{e55}\end{eqnarray}

According to Ref.\cite{GKW79},the strength of each current 
could be localized around some inelastic scattering center $x_i$, 
such that 
 
\begin{equation}
j_i(x)=j(x-x_i) \; \; \; ;  \; \; \; |A_i|^2 =\rho(x_i)
\; \;.\end{equation}

The current $j(x-x_i)$ is considered to be peaked 
around $x_i$, and could be characterized by the 
size scale of the wave packet; $\rho(x_i)$ is the 
source distribution function of the emitter. 
Naturally, we are now considering a simplified picture, in which 
phase-space correlations are absent. 
Then Eq. (\ref{e55})
can be further simplified as
\begin{eqnarray}
C_2(p_1,p_2)&=&1+\frac{\sum_{i,j} \rho(x_i)\rho(x_j)cos[(p_1-p_2)(x_i-x_j)]}
{\sum_i \rho(x_i) \sum_k \rho(x_k)}
\nonumber\\
&&
-
\frac{\sum_i \rho(x_i)\rho(x_i) }
{\sum_i \rho(x_i) \sum_k \rho(x_k)}. 
\label{c2rho}\end{eqnarray}

The last term in Eq. (\ref{p2semi}), (\ref{e55}) and (\ref{c2rho}) 
discounts the contribution corresponding to emitting two pions 
from the {\it same} source. In the case of very large volumes
this type of contribution is usually considered to be small when compared 
to the emission from separate sources. 
Consequently, in cases where this term $\cal{O}$ $(1/V$) can be 
neglected\cite{GKW79,Heinz96},  
we recover the well-known two-pion interferometry formula 
\begin{equation}
C_2({\bf p_1,p_2})=1+\int \rho(x)\rho(y) cos\left[(p_1-p_2)(x-y)\right]
 dx dy
.\label{e58}\end{equation} 

In Ref.\cite{pgg}, a general semi-classical approach to two-pion 
interferometry was used, in which a Gaussian wave packet spread was 
allowed for incorporating minimal effects due to the uncertainty principle. 
As pointed out in the above reference, Eq. (\ref{e58}) 
corresponds to approaching the classical regime, i.e., 
it would be valid only when the wave packet size is negligible, 
which would also be equivalent to consider system sizes much bigger 
than the wave packet size. Being so, 
the above derivation would be considered as a good approximation only in 
cases where the source size is large, as in heavy-ion collisions. However, 
we should be cautious when using it in $e^+e^-$ collisions as, in that 
case, the source radius is small and the third term in Eq. (\ref{c2rho}) 
may not be negligible. Besides, 
the chaotic source ansatz is also questionable there. 
Just to emphasize this point, let us naively 
consider a fictitious source 
of 0 fm size, i.e., $\rho(x) = \delta(x)$. Then, from Eq. (\ref{e58}), 
we would get 
\begin{eqnarray}
C_2({\bf p_1,p_2})&=& 2,
\end{eqnarray} 
and, for the chaoticity parameter 
\begin{equation}
\lambda = C_2({\bf p,p})-1=1. 
\end{equation}

Naturally, we cannot think of a ``zero size''  source as being chaotic. 
Actually, as it has been stated in Ref.\cite{GKW79}, and 
illustrated above, the chaotic ansatz is only correct when the source 
size $V$ is much larger than the size of the wave packet. In the 
fictitious source case above, if we do not neglect the 
third term in Eq. (\ref{e55}), we would obtain 
\begin{equation}
C_2({\bf p_1,p_2})=1,
\end{equation}
which is the correct two-pion interferometry result for the above 
model\cite{expl}. 

For confronting 
the role of the correction term in Eq. (\ref{e55}) 
and Eq. (\ref{c2rho}) with the correlation 
function estimated by using Eq. (\ref{e58}) we choose, 
for simplicity, $A_i=1/\sqrt{V}$ and 
$j_i(x)=1/\sqrt{V_{\sigma}}$, where $V$ is the total volume of the 
source and $V_{\sigma}$ is the volume of the emitter, which is of the 
order of the wave packet size. In Figure 6 we show the corresponding results.  
We see  that, for $V=(5fm)^3$ and $V_{\sigma}=(1fm)^3$, 
the third term is much smaller then the second, and could be 
neglected, while for $V=(2fm)^3$ and $V_{\sigma}=(1fm)^3$, 
its correction is substantial. Clearly then, these results 
depend strongly on the wave packet size: the smaller it is with 
respect to the system size, the better is the approximation 
represented by Eq. (\ref{e58}). 
From the results in Figure 6, it seems that, if the wave packet 
size is about $1fm$, we could use the conventional pion interferometry 
formula in Eq. (\ref{e58}) for analyzing pion interferometry in heavy-ion 
collisions. However, 
we could not use it to analyze $e^+e^-$ collisions, since in that 
case the source radius is of the same order as the wave packet size, and 
the contribution the third term is non-negligible.
It is interesting to notice that the above derivation is 
equivalent to the one in Ref.\cite{GKW79,Heinz96} 
where a density matrix formulation was also used. 
Appendices I and II contain, respectively, further discussion 
regarding a density matrix formulation 
leading to an equivalent of Eq. (\ref{e55}), and a simple 
unified form for the two formulations. 

\section{Conclusions}

In this paper, we derive the two-pion correlation function 
by adopting a different density matrix, as given in Eq.(\ref{rho}). 
The finite volume effects on the pion spectrum were then  
studied in Figures 1, 2 and 3, leading to similar results as 
in Ref.\cite{MW93,MW95,AS97}.  
We found that the small momentum 
region is depleted with respect to the Bose-Einstein distribution. 
The effects on two equally charged pion correlation function were also 
analyzed. The results in Figures 4 and 5 show that the correlation 
function shrinks for increasing average pair momentum and for increasing 
size of the emitting source, respectively, corresponding to an increase 
of its inverse width. 
The first result reflects a strong sensitivity to the dynamical matrix, 
through the Bose-Einstein weight factor, as discussed in Section III.A.  
We also discussed the effects of a diffuse boundary on the spectrum and 
correlation function by considering a smooth decrease to zero of the 
wave function as $R$ goes to infinity. 

We compared as well 
the results obtained by means of Eq. (\ref{pi2}) with 
those estimated by using Eq. ({\ref{e55}) or ({\ref{c2rho}), 
showing that they may differ significantly when small volumes are 
considered. For instance, from Figures 4 and 5, 
and as a result of Eq. (\ref{pi2}), we see that the boundary affects 
the single- and the double-inclusive distributions in a consistent way, 
as the intercept of the correlation function remain 
unchanged when altering the system size. Nevertheless, when 
looking at the curves generated by using Eq. ({\ref{e55}) 
in Figure 6, we see that the intercept of $C(q)$ drops as the size of the 
system is reduced, leading to unphysical results. 
The relation of the system size to the wave packet size 
is the key to understand this behavior: when this last one cannot  
be considered much smaller than the first one, the additional term in 
Eq.(\ref{e55}) or Eq.(\ref{c2rho}) would give a non-negligible contribution. 
Otherwise, Eq.(\ref{e55}) or Eq.(\ref{c2rho}) would approximate 
the conventional HBT formulation, represented by 
Eq.(\ref{e58}), which  
was derived under the condition that the 
system size is much larger than the wave packet size. 
Nevertheless, we should notice that this assumption was 
{\it not} necessary for obtaining Eq. (\ref{pi2}), since this 
was derived strictly within the quantal realm. 
It is interesting to remark that, in the simplistic case of a Gaussian 
breakup distribution, considering  wave packets with non-negligible widths, 
Eq. (61) of Ref.\cite{pgg} showed that the inverse width is 
enlarged in direct proportion to the wave packet size, 
$R^2_\Delta = R^2 + \Delta x^2 - \frac{1}{4(P^2 + \Delta p^2)}$. 
In this expression, $R$ is the Gaussian width in space-time,  
$P$ is the width of the Gaussian distribution in the momentum space; 
$\Delta x$ and $\Delta p$ are the corresponding wave packet spread. 
For $R \sim 2$ fm, $P \sim 0.14$ GeV/c, and 
minimum packets with $\Delta x = 1$ fm, 
the corresponding inverse width would be $R_\Delta \approx 2.16$ fm, 
an increase of about eight percent. This very rough estimate 
seems to be of the same order of the decrease in width 
(i.e., increase in the apparent radius) seen in Figure 6. 
Finally, in Appendix I we show that, by adopting another 
density matrix instead of the one proposed in Eq. (\ref{rho}), we can 
derive an interferometry result which is similar to Eq. (\ref{e55}). 
And in Appendix II, we suggest a simplified way of unifying these two 
formulations, i.e., of recovering each of them by means of a parameter choice.

\bigskip\bigskip
\begin{center}
{\bf Acknowledgements}
\end{center}
We thank C.Y. Wong for elucidating discussions. We would also like to 
express our gratitude to M. Gyulassy and Yu. Sinyukov 
for several helpful discussions.  This work was partially supported 
by the Funda\c{c}\~ao de Amparo \`a Pesquisa do Estado de S\~ao Paulo 
(FAPESP), Brazil (Proc. N. 1998/05340-2 and 1998/2249-4).

\vfill
\newpage

\section{Appendix I}

In what follows, we show that, by considering the following 
density matrix 

\begin{equation}
\rho=\prod_{\lambda}\rho_{\lambda}
\end{equation}
instead of the one proposed in Eq. (\ref{rho}), we can 
also derive an interferometry result which is similar to Eq. (\ref{e55}), 
where 

\begin{equation}
\rho_{\lambda}=\sum_{n_{\lambda}=0}^{\infty} 
\frac{(a_{\lambda}^{\dagger})^{n_{\lambda}}}
{\sqrt{n_{\lambda}!}}|0\rangle 
\frac{\exp(-n_{\lambda}\cdot (E_{\lambda}-\mu_{\lambda})/T)}{n_{\lambda}!}
\langle 0| \frac{(a_{\lambda})^{n_{\lambda}}}{\sqrt{n_{\lambda}!}}.
\end{equation}

The corresponding pion multiplicity distribution is given by

\begin{eqnarray}
P_{n_{\lambda}}&=&\frac{\langle n_{\lambda} \rangle^{n_{\lambda}}}{n_{\lambda}!}
\exp(-\langle n_{\lambda} \rangle),  
\nonumber\\
\langle n_{\lambda} \rangle&=& N_{\lambda}= \exp(-(E_{\lambda}-\mu)/T),
\nonumber
\end{eqnarray}

and the single particle spectrum is then 

\begin{equation}
P_1({\bf p})=\sum_{\lambda}N_{\lambda}|\tilde{\psi}_{\lambda}({\bf p})|^2.
\end{equation}

In the limit $V \rightarrow \infty$, we obtain the {\it Boltzmann distribution} 
instead of {\it Bose-Einstein distribution} derived previously in Section III. 

Similarly, we can derive the two-pion correlation function as

\begin{eqnarray}
C_2(p_1,p_2)&=&1+\frac{|\sum_{\lambda} N_{\lambda}\psi_{\lambda}^*(p_1)
\psi_{\lambda}(p_2)|^2}
{\sum_{\lambda} N_{\lambda}|\psi_{\lambda}(p_1)|^2
\sum_{\lambda} N_{\lambda}|\psi_{\lambda}(p_2)|^2}
\nonumber\\
&&
-\frac{\sum_{\lambda} N_{\lambda}^2|\psi_{\lambda}
(p_1)\psi_{\lambda}(p_2)|^2 }
{\sum_{\lambda} N_{\lambda}|\psi_{\lambda}(p_1)|^2
\sum_{\lambda} N_{\lambda}|\psi_{\lambda}(p_2)|^2}
\; \;. \label{pi211}\end{eqnarray}

The major difference between this equation and Eq. (\ref{pi2}) 
has its origin in the fact that   
now we have 

\begin{equation}
\langle a^{\dagger}_{\lambda}a^{\dagger}_{\lambda}a_{\lambda}a_{\lambda}
 \rangle = \langle a^{\dagger}_{\lambda}a_{\lambda}\rangle^2
\label{ceboltz}\end{equation}
instead of Eq. ({\ref{BE1}}), in section II. This comes from the fact 
that, in Eq. (\ref{ceboltz}), we are dealing with classical (i.e., 
distinguishable) particles. In Appendix II, we show a way to unify the 
formulation leading to Eq. (\ref{pi2}) and Eq. (\ref{pi211}).

\vfill
\newpage 

\section{Appendix II}

If we assume that a single pion state  
could be described by the quantum number $\lambda$, then 
the single particle spectrum could be expressed as

\begin{equation}
P_1(p)=\sum_{\lambda} \omega_{\lambda}|\tilde{\psi}_{\lambda}({\bf p})|^2.
\end{equation}

Here $\omega_{\lambda}$ is the occupation probability of the single-particle 
state $\lambda$. In the two particle distribution case, the pions could be 
described by two different quantum numbers, $\lambda_1$ and $\lambda_2$. 
By imposing the symmetrization required by the 
Bose-Einstein statistics, the two-pion wave function could be 
written as

\begin{eqnarray}
%\begin{equation}
\psi_{\lambda_1,\lambda_2}({\bf p_1,p_2})&=&\frac{1}{\sqrt{2}}
[\tilde{\psi}_{\lambda_1}({\bf p_1})
\tilde{\psi}_{\lambda_2}({\bf p_2})+ 
\nonumber\\
&& \; \; \; \;\;  \;\;  \tilde{\psi}_{\lambda_1}({\bf p_2})
\tilde{\psi}_{\lambda_2}({\bf p_1})].
\end{eqnarray}
%\end{equation}

If the two-pions are in the same state, then we would 
have 

\begin{equation}
\psi_{\lambda,\lambda}({\bf p_1,p_2})=
A\cdot \tilde{\psi}_{\lambda}({\bf p_1})\tilde{\psi}_{\lambda}({\bf p_2}). 
\end{equation}

If the multiplicity distribution follows the geometry distribution, then
$A=\sqrt{2}$. If, however, multiplicity distribution has Poisson form
then $A=1$ . Being so, the two-pion distribution could  be 
expressed in terms of $A$ as

\begin{eqnarray}
P_2({\bf p_1,p_2})&=&\sum_{\lambda_1,\lambda_2,\lambda_1 \ne \lambda_2}
\omega_{\lambda_1}\omega_{\lambda_2}
\nonumber\\
&&
\frac{1}{2} \left [ \frac{}{}
|\tilde{\psi}_{\lambda_1}({\bf p_1})|^2
|\tilde{\psi}_{\lambda_2}({\bf p_2})|^2 +
\right.
\nonumber\\
&&
|\tilde{\psi}_{\lambda_2}({\bf p_1})|^2
|\tilde{\psi}_{\lambda_1}({\bf p_2})|^2 +
\nonumber\\
&&
\tilde{\psi}_{\lambda_1}^*({\bf p_1})\tilde{\psi}_{\lambda_1}({\bf p_2})
\tilde{\psi}_{\lambda_2}^*({\bf p_2})\tilde{\psi}_{\lambda_2}({\bf p_1})
\nonumber\\
&&
\left.
\tilde{\psi}_{\lambda_1}^*({\bf p_2})\tilde{\psi}_{\lambda_1}({\bf p_1})
\tilde{\psi}_{\lambda_2}^*({\bf p_1})\tilde{\psi}_{\lambda_2}({\bf p_2})\right ]\frac{}{}
\nonumber\\
&&+ |A|^2 \sum_{\lambda}\omega_{\lambda}^2
|\tilde{\psi}_{\lambda}({\bf p_1})\tilde{\psi}_{\lambda}({\bf p_2})|^2
\nonumber\\
&=&P_1({\bf p_1})P_1({\bf p_2})+
\nonumber\\
&&
|\sum_{\lambda}\omega_{\lambda}\tilde{\psi}^*_{\lambda}({\bf p_1})
\tilde{\psi}_{\lambda}({\bf p_2})|^2
\nonumber\\
&&
-2\sum_{\lambda}\omega_{\lambda}^2|\tilde{\psi}_{\lambda}({\bf p_1})
\tilde{\psi}_{\lambda}({\bf p_2})|^2
\nonumber\\
&&+ |A|^2 \sum_{\lambda}\omega_{\lambda}^2
|\tilde{\psi}_{\lambda}({\bf p_1})\tilde{\psi}_{\lambda}({\bf p_2})|^2
\end{eqnarray}

	Consequently, for different choices of A, i.e., $A=\sqrt{2}$ 
or $A=1$ discussed above, 
we could recover the results in Eq. (\ref{pi2}) or Eq. (\ref{pi211}), 
respectively.

%%%%%%%%%%%%%%%%%%%%%%%%%%%%%%  FIGURES  %%%%%%%%%%%%%%%%%%%%%%%%%%%%%%

\newpage
\vskip 5cm
\begin{figure}[h]\epsfxsize=8cm
\centerline{\epsfbox{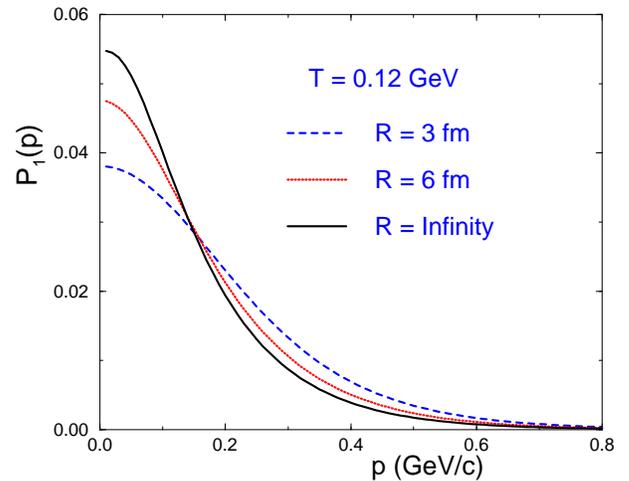}}
\caption{\it Normalized spectrum (in arbitrary units) 
vs. momentum $|${\bf p}$|$ (in GeV/c). 
The input temperature is $T=0.12$ GeV and the chemical potential is 
$\mu=0$. The solid line
corresponds to the Bose-Einstein distribution, i.e., to the limit   
$R \rightarrow \infty$. The dotted and dashed lines correspond, respectively,  
to the $R=6$ fm and $R=3$ fm cases.}
\end{figure}

\begin{figure}[h]\epsfxsize=8cm
\centerline{\epsfbox{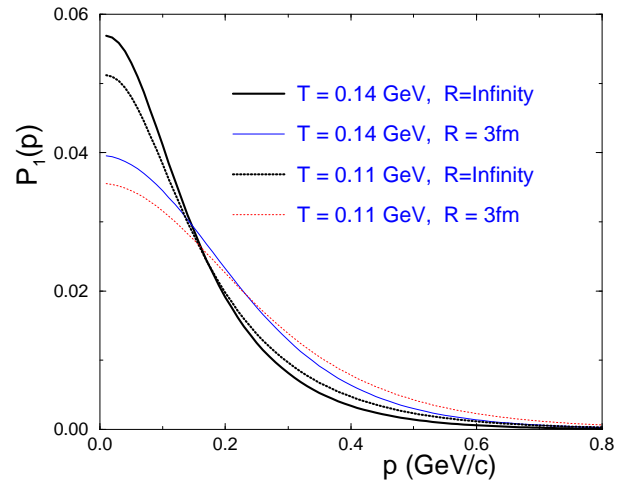}}
\caption{\it Normalized single-particle distribution (in arbitrary units) 
vs. momentum $|${\bf p}$|$.
 The input parameters are $\mu=0$,  and the radius in the finite case, 
 $R=3$ fm. The curves are shown for two values of the freeze-out temperature:  
$T=0.14$ GeV (solid) and  $T=0.11$ GeV (dotted)
and  compared with the corresponding ones in the 
$R \rightarrow \infty$ limit. }
\end{figure}

\begin{figure}[h]\epsfxsize=8cm
\centerline{\epsfbox{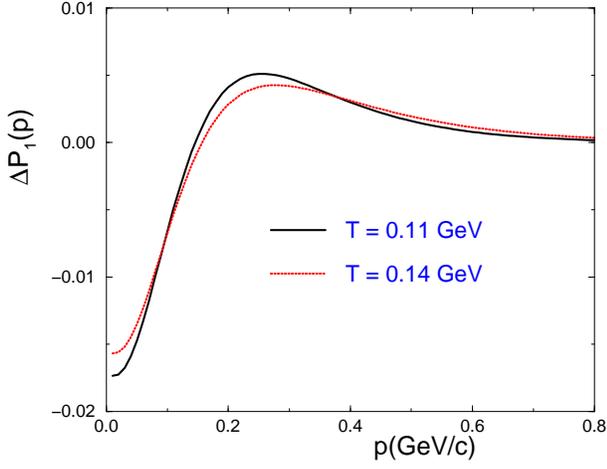}}
\caption{\it Difference in the momentum distribution (in arbitrary units), 
$\Delta P_1({\bf p})= P_1({\bf p})|_{R=3fm}-P_1({\bf p})|_{R=\infty}$, 
vs. $|${\bf p}$|$, of curves with  
$R=3$ fm and the corresponding ones in the  
$R \rightarrow \infty$ limit, 
for two values of the freeze-out 
temperature $T$, as indicated in the plot, with $\mu=0$.  }
\end{figure}

\begin{figure}[h]\epsfxsize=8cm
\centerline{\epsfbox{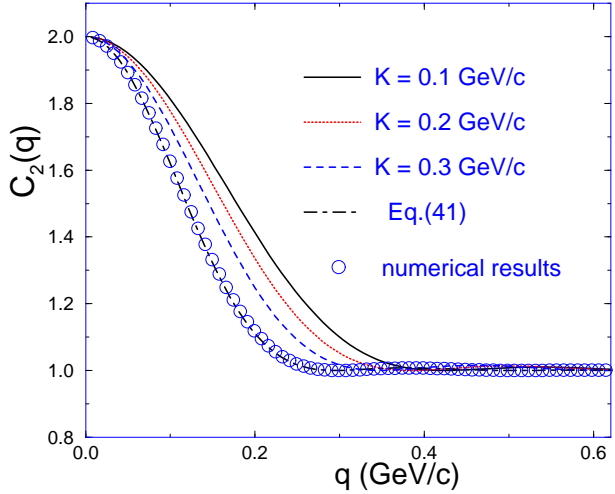}}
\caption{\it Two-pion correlation function, $C_2(q)$, is shown versus the 
momentum difference $|${\bf q}$|$=$|$${\bf p_1-p_2}$$|$.
The input parameters are $T=0.12$ GeV, $R= 3$ fm, and $\mu=0$. The solid,
dotted, and dashed lines correspond to the average pair 
momentum values $K$= 0.1 GeV/c, 0.2 GeV/c, and 0.3 GeV/c, respectively.
The circles refer to numerical results similar to the previous ones but 
with unity weight factor, $N_\lambda = 1$, in Eq. (\ref{pi2}). 
The dot-dashed line corresponds to the analytical result in Eq. (41).}
\end{figure}

\begin{figure}[h]\epsfxsize=8cm
\centerline{\epsfbox{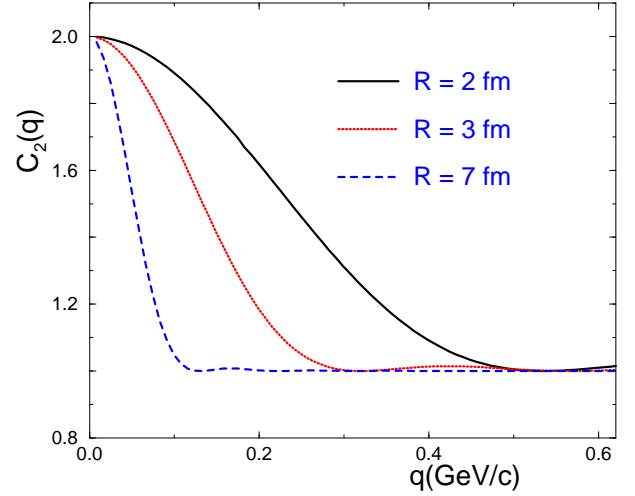}}
\caption{\it The correlation function, $C_2(q)$, is 
plotted as a function of the momentum difference, 
$|${\bf q}$|$=$|$${\bf p_1-p_2}$$|$. The input parameters are 
the temperature  
$T=0.12$ GeV, the average pair momentum $K = 0.4$ GeV/c, and $\mu=0$. 
The solid,
dotted, and dashed lines correspond, respectively, to the sphere radius 
$R$ = $2$ fm, $3$ fm, and $7$ fm. }
\end{figure}

\begin{figure}[h]\epsfxsize=8cm
\centerline{\epsfbox{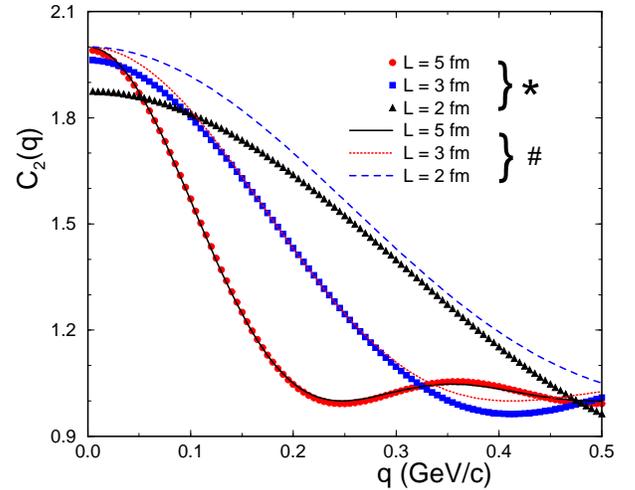}}
\caption{\it Two-pion cor\-re\-la\-tion func\-tion, $C_2(q)$,  is shown 
versus $|${\bf q}$|$=$|$${\bf p_1-p_2}$$|$. 
For helping visualization the curves are separated by brackets in two groups. 
In the first, signaled by ($\star$), the curves were obtained with the help 
of Eq. (\ref{e55}). The other group,  signaled by ($\#$), 
corresponds to curves obtained by means of Eq. (\ref{e58}). 
The values adopted for the finite system sizes, i.e., 
$L$= $2$ fm, $3$ fm, 
and $5$ fm, are shown in the plot. }
\end{figure}


\begin{thebibliography}{29}
\bibitem{expl}
It was shown in Ref.\cite{GKW79} that, since for a point-like source 
we have only one emitter, the corresponding pion multiplicity 
distribution is Poisson and the two-pion correlation function is 1.
\end{thebibliography}
\end {document}